\documentclass[seceq]{ptptex}

\usepackage{graphicx}

\newcommand{\del}{\partial}

\newcommand{\dd}{{\rm d}}
\newcommand{\ee}{{\rm e}}

\title{
Quark Number Susceptibility with Finite Quark Mass \\
in Holographic QCD
\footnote{{\tt arXiv:1012.2667[hep-ph], APCTP-Pre2010-008, 
CAS-KITPC/ITP-211}}
}

\author{
Kyung-il \textsc{Kim}$^1$\footnote{E-mail: hellmare@yonsei.ac.kr},
Youngman \textsc{Kim}$^{2,3}$\footnote{E-mail: ykim@apctp.org},
Shingo \textsc{Takeuchi}$^4$\footnote{E-mail: shingo@itp.ac.cn}
\\
and
Takuya \textsc{Tsukioka}$^2$\footnote{E-mail: tsukioka@apctp.org}
}

\inst{
$^1$Institute of Physics and Applied Physics,
Yonsei University, Seoul 120-749, Korea
\\
$^2$Asia Pacific Center for Theoretical Physics,
Pohang, Gyeongbuk 790-784, Korea
\\
$^3$Department of Physics, Pohang University of Science and Technology,
\\
Pohang, Gyeongbuk 790-784, Korea
\\
$^4$Kavli Institute for Theoretical Physics China,
Institute of Theoretical Physics, \\
Chinese Academy of Sciences, Beijing 100190, China \\
}

\abst{
We study the effect of a finite quark mass on the quark number
susceptibility in the framework of holographic QCD.
We work in a bottom-up model with a deformed AdS black hole
and D3/D7 model to calculate the quark number susceptibility
at finite temperature with/without a finite quark chemical potential.
As expected the finite quark mass suppresses the quark number
susceptibility.
We find that at high temperatures $T\ge 600$ MeV the quark
number susceptibility of light quarks and heavy quarks
are almost equal in the bottom-up model.
This indicates that the heavy quark like charm contribution
to thermodynamics of a QCD-like system may start to become
significant at temperatures $T\sim 600$ MeV.
In D3/D7 model, we focus on the competition between
the quark chemical potential, which enhances the quark number
susceptibility, and the quark mass that suppresses the susceptibility.
We observe that depending on the relative values of the quark mass
and the quark chemical potential, the quark number susceptibility
shows a diverging or converging behavior.
We also calculate the chiral susceptibility in D3/D7 model to support
the observation made with the quark number susceptibility.
}

\begin{document}

\maketitle

\section{Introduction}

Fluctuations of conserved charges such as baryon number,
electric charge, and strangeness are generally considered
as useful probes for the structure of a thermal medium,
quark gluon plasma (QGP), produced in heavy ion collisions.
Quark number fluctuations are basic observables which can be
obtained by taking derivatives of the grand canonical potential
with respect to the quark chemical potential.
The existence of a peak in the second order quark number
fluctuation, or quark number susceptibility, near $T_c$ is
confirmed by recent lattice QCD calculations based on the
Taylor expansion with respect to the quark (or baryon) chemical
potential~\cite{Alltonetal, Ejiri}.
This implies the existence of the critical end point (CEP),
at which the first order phase
transition terminates in the $(\mu_q, T)$ plane of the QCD phase diagram
(see \cite{CEP} for a review).
There have been many studies to calculate the quark
number susceptibility and higher order quark number fluctuations in
various model
studies~\cite{Mc87, Gottlieb:1987ac, Kunihiro:91, CMT, HI, HKRS,
Sasaki, HM2010} and lattice
simulations~\cite{GLTRS, GLTRS88, GPS, GKetal, EKR}.
In \cite{jkls, KMSTT, SZ}, the quark number susceptibility at
finite temperature with finite chemical potential is studied
in a holographic QCD model.

The AdS/CFT correspondence \cite{Maldacena,gkp,w} is a powerful
tool to study strongly coupled gauge theories.
Using this correspondence, we can obtain
physical quantities in gauge theories from gravity side.
We can also discuss finite temperature systems from
this gauge/gravity correspondence.
In the hydrodynamic limit i.e.\ low frequency limit,
we can calculate correlation functions~\cite{Son:2002sd, pss}.
The most fruitful hydrodynamic result is the observation of
the ``universal'' bound of the ratio of the shear viscosity to entropy
density~\cite{kss}.
Although the correspondence between QCD and gravity theory is not known,
we can obtain much insights by using this correspondence.

In this paper we study the effect of a finite quark
mass on the quark number susceptibility in holographic QCD.
We consider light, strange and heavy (charm) quarks.
In the previous studies \cite{jkls, KMSTT}, the mass of quark
is zero, and so the quark number susceptibility with two massless
flavor was calculated in a QCD-like system.
We now extend the previous work by considering no-zero quark masses.
Our basic motivation for the charm quark is an observation made
in perturbative QCD analysis on the role of quark mass in QCD
thermodynamics.  In~\cite{LS}, it is claimed that unexpectedly,
the charm quark plays some role in the QGP at relatively low
temperatures $T\sim 350$ MeV.
Since this analysis is based on the perturbation, it should be
important to check the claim in a non-perturbative study.
To explore the  effect of a finite quark mass at finite temperature
and/or density, we work with two models.
The first one is a bottom-up type model based on the deformed AdS
black hole due to finite quark mass obtained in \cite{KMS}.
Then we consider D3/D7 model with a finite chemical potential
studied in \cite{D3D7density, kmmmt}.

This paper is organized as follows:
After introducing the quark number susceptibility,
we calculate it in the deformed AdS black hole background
in Section 2.
The hydrodynamic and thermodynamic analysis are given in the section.
In Section 3, we discuss the quark number susceptibility
in D3/D7 system.
We summarize the result in the final section.
The procedure to obtain the retarded Green function
is briefly given in the Appendix A.
In Appendix \ref{AppB}, we consider the chiral susceptibility
in D3/D7 model, though the chiral symmetry in D3/D7 is only
$U(1)_{\rm A}$ not that of QCD,
$SU(N_f)_{\rm L}\times SU(N_f)_{\rm R}$.

\section{Quark number susceptibility in deformed AdS black hole}

The quark number susceptibility was proposed as
a probe of the QCD chiral phase transition at zero chemical
potential~\cite{Mc87, Gottlieb:1987ac},
\begin{equation}
\chi_q=\frac{\del n_q}{\del\mu}.
\label{def_qns}
\end{equation}
Later it has been shown that the quark number susceptibility
can be rewritten in terms of the retarded Green
function through the fluctuation-dissipation
theorem~\cite{Kunihiro:91},
\begin{equation}
\chi_q
=-\lim_{k\to 0} {\mbox{Re}}\Big(G_{t\ t}(\omega=0, k)\Big),
\label{suss}
\end{equation}
where $G_{\mu \ \nu}(\omega, k)$ is the retarded Green function
of the vector-vector correlations.

Now, we calculate the quark number susceptibility
in the deformed AdS black hole background.
We do this in two different ways that should be equivalent each other.
We first use the equation (\ref{suss}) to conduct hydrodynamic
analysis and then calculate the quark number susceptibility
using the thermodynamic relation in the equation (\ref{def_qns}).

The background we shall discuss is
AdS black hole with back-reactions coming from flavors~\cite{KMS},
\begin{equation}
(\dd s)^2
=
\ee^{-2K(z)}\Big( - f(z)(\dd t)^2
+ (\dd\vec{x})^2 \Big)
+ \frac{(\dd z)^2}{z^2f(z)},
\label{metric_dads_01}
\end{equation}
with
\begin{subequations}
\begin{eqnarray}
K(z) &=& \log z + \frac{1}{4}\left(\frac{z}{z_Q} \right)^2,
\label{K(z)}
\\
f(z) &=& 1-2\left( \frac{z_Q}{z_h} \right)^4
\left\{ 1-\left( 1- \left( \frac{z}{z_Q} \right)^2 \right)
         \ee^{(z/z_Q)^2} \right\}.
\label{f(z)}
\end{eqnarray}
\end{subequations}

\noindent
The parameter $z_Q$ is defined as
\begin{equation}
z_Q^2=\frac{6}{\kappa^2 N_f M_q^2}.
\label{zq}
\end{equation}
$M_q$ represents the quark mass and $\kappa^2=8\pi G_5$ with
$1/G_5=32N_c^2/(\pi L^3)$ where $L$ is the AdS radius.
In \cite{KMS}
there were two deformations to be discussed, deformed AdS and
deformed AdS black hole.
In the present work we only consider one of them, i.e.\ the deformed AdS
black hole background.
Each of these two backgrounds is reasonable with given
potentials\footnote{
In \cite{KMS} the authors discussed
the Hawking-Page type transition~\cite{hp, hpw} which is the dual to the
confinement/deconfinement transition using these two deformed backgrounds.
However, since the bulk potentials for each background are not the same,
the Hawking-Page type analysis
in \cite{KMS} needs to be improved.
}.
Note that in \cite{KMS} $N_f$ counts the number of
massive quarks only.
For instance, if we are interested in a 2+1 flavor system,
we take $N_f=1$ in the equation (\ref{zq}),
neglecting the back-reaction from the two flavors
whose masses are zero or very small.
Taking the limit $M_q\to 0$,
the metric (\ref{metric_dads_01}) is reduced to
the AdS black hole
$$
(\dd s)^2=\frac{1}{z^2}
\Big(-f(z)(\dd t)^2+(\dd\vec{x})^2\Big)
+\frac{(\dd z)^2}{z^2f(z)},
$$
with
$$
f(z)=1-\bigg(\frac{z}{z_h}\bigg)^4,
$$
where $z_h$ implies the location of horizon of the
original AdS black hole.
There exists a horizon $z_H$ which gives zero of the function
(\ref{f(z)})\footnote{
We may refer the position of horizon as
$$
z_H=z_Q\sqrt{1+\mbox{ProductLog}\bigg(\frac{(z_h/z_Q)^4-2}{2\ee}\bigg)},
$$
where ProductLog($a$) stands for a solution for $x$ in the equation
$x\ee^x=a$.
}.
Then the Hawking temperature is given by
\begin{equation}
T=\frac{1}{\pi z_H}\Big(\frac{z_H}{z_h}\Big)^4
\ee^{\frac{3}{4}\big(\frac{z_H}{z_Q}\big)^2}.
\label{ht}
\end{equation}
We note that for small $M_q$, $z_H$ becomes $z_h$. 
Using the background (\ref{metric_dads_01}), 
we assume $M_q$ is not large within the small mass deformation. 
Since the background (\ref{metric_dads_01}) is constructed from the 
bottom-up type approach, 
we could not find any restrictions for the finite value of $M_q$.  

Let us now consider 5D $U(1)$ gauge field which is dual to
4D quark number current on the
background (\ref{metric_dads_01}).
We then calculate the quark number susceptibility
by using the AdS/CFT correspondence.
Since in this background there is no background charge which
could be interpreted as the chemical potential in the boundary theory,
we here discuss the mass dependence of the quark number susceptibility.

We start by introducing a dimensionless coordinate
$u\equiv (z/z_H)^2$ which is normalized by the horizon.
In this coordinate system, the horizon and the boundary are located at
$u=1$ and $u=0$, respectively.
The metric (\ref{metric_dads_01})
is rewritten as
\begin{equation}
(\dd s)^2
=\ee^{-2K(u)}\Big(-f(u)(\dd t)^2+(\dd\vec{x})^2\Big)
+\frac{(\dd u)^2}{4u^2f(u)},
\label{metric_dads_02}
\end{equation}
with
\begin{subequations}
\begin{eqnarray}
K(u)
&=&
\frac{1}{2}\log(z_H^2u)+\frac{1}{4}\Big(\frac{z_H}{z_Q}\Big)^2u,
\label{K(u)}
\\
f(u)
&=&
1-2\left( \frac{z_Q}{z_h} \right)^4
\left\{ 1-\left( 1- \left( \frac{z_H}{z_Q} \right)^2\!u \right)
         \ee^{\big(\frac{z_H}{z_Q}\big)^2\!u} \right\}
\nonumber
\\
&=&
\frac{1}{1-\bigg(1-\Big(\frac{z_H}{z_Q}\Big)^2\bigg)
\ \ee^{\big(\frac{z_H}{z_Q}\big)^2}}
\Bigg\{
\bigg(1-\Big(\frac{z_H}{z_Q}\Big)^2\!u\bigg)
\ \ee^{\big(\frac{z_H}{z_Q}\big)^2\!u}
\nonumber
\\
&&
\hspace*{46mm}
-\bigg(1-\Big(\frac{z_H}{z_Q}\Big)^2\bigg)
\ \ee^{\big(\frac{z_H}{z_Q}\big)^2}
\Bigg\}.
\label{f(u)}
\end{eqnarray}
\end{subequations}

The action for the $U(1)$ gauge field is
\begin{equation}
S=-\frac{1}{4g_5^2}\!\int\!\dd^5x
\sqrt{-g}F_{mn}F^{mn},
\end{equation}
where $g_5$ is the 5D gauge coupling constant.
We shall work in $A_u(x)=0$ gauge and use the Fourier decomposition
\begin{equation}
A_\mu(t, z, u)=\!\int\!\frac{\dd^4k}{(2\pi)^4}
\ \ee^{-i\omega t+ikz}A_\mu(\omega, k, u),
\end{equation}
where we choose the spacial momenta which are along the $z$-direction.
Variations of the action with respect to
$A_t(u)$ and $A_u(u)$ give equations of motion
\begin{subequations}
\begin{eqnarray}
0
&=&
\Big(\ee^{-\frac{1}{2}\big(\frac{z_H}{z_Q}\big)^2\!u}A'_t(u)\Big)'
-\frac{z_H^2}{4uf(u)}
\Big(k^2A_t(u)+\omega k A_z(u)\Big), \quad
\label{eom_at}
\\
0
&=&
\omega A_t'(u)+kf(u)A_z'(u),
\label{eom_au}
\end{eqnarray}
\end{subequations}

\noindent
where the prime stands for the derivative with respect to $u$.
The equation of motion for $A_z(u)$ can be derived from the equations
(\ref{eom_at}) and (\ref{eom_au}).
For $A_x(u)$ and $A_y(u)$, one can obtain decoupled equations
of motion.
Since we are interested in the time-time component of the retarded
Green function to calculate the quark number susceptibility (\ref{suss}),
we will not consider $A_x(u)$ and $A_y(u)$ hereafter.

From the equations (\ref{eom_at}) and (\ref{eom_au}),
we obtain an equation for $A_t(u)$,
\begin{eqnarray}
0
&=&
\Big(uf(u)\ \ee^{-\big(\frac{z_H}{z_Q}\big)^2\!u}A_t''(u)\Big)'
\nonumber
\\
&-&\Bigg\{
\frac{1}{2}\Big(\frac{z_H}{z_Q}\Big)^2
\ee^{-\frac{1}{2}\big(\frac{z_H}{z_Q}\big)^2\!u}
\Big(uf(u)\ \ee^{-\frac{1}{2}\big(\frac{z_H}{z_Q}\big)^2\!u}\Big)'
-\frac{z_H^2}{4}\ \ee^{-\frac{1}{2}\big(\frac{z_H}{z_Q}\big)^2\!u}
\Big(\frac{\omega^2}{f(u)}-k^2\Big)
\Bigg\}A_t'(u).
\nonumber
\\
\label{eom_at_0}
\end{eqnarray}
We can make the equation more convenient form by changing the
variable,
\begin{equation}
A'_t(u)=\ee^{\frac{1}{2}\big(\frac{z_H}{z_Q}\big)^2\!u}X(u).
\end{equation}
Then the equation (\ref{eom_at_0}) can be rewritten
as
\begin{equation}
0=\Big(uf(u)X'(u)\Big)'
+\frac{z_H^2}{4}\ \ee^{\frac{1}{2}\big(\frac{z_H}{z_Q}\big)^2\!u}
\bigg(\frac{\omega^2}{f(u)}-k^2\bigg)X(u).
\label{eom_at_02}
\end{equation}

Let us proceed to solve the equation of motion (\ref{eom_at_02}).
This is an ordinary second order differential equation with
a regular singularity at the horizon $u=1$.
Writing a solution as $X(u)=(1-u)^\nu F(u)$ where
$F(u)$ is a regular function at the horizon, we fix the
constant $\nu$ by imposing the incoming wave condition,
\begin{equation}
\nu=-i\frac{\omega}{4\pi T},
\end{equation}
where $T$ is the Hawking temperature (\ref{ht}).

Now we shall solve the equation of motion in the hydrodynamic regime
i.e.\ small $\omega$ and $k$ compared with the temperature $T$.
In this regime, we could expand the function $F(u)$ as
\begin{equation}
F(u)=F_0(u)+\omega F_\omega(u)+k^2 F_{k^2}(u)
+{\cal O}(\omega^2, \omega k^2).
\end{equation}
Since we are interested in the quark number susceptibility defined
by (\ref{suss}), we only need solutions for $F_0(u)$ and $F_{k^2}(u)$.
Relevant equations are read off as
\begin{subequations}
\begin{eqnarray}
0
&=&
\Big(uf(u)F'_0(u)\Big)',
\label{eom_f0}
\\
0
&=&
\Big(uf(u)F'_{k^2}(u)\Big)'-\frac{z^2_H}{4}\
\ee^{\frac{1}{2}\big(\frac{z_H}{z_Q}\big)^2\!u}
F_0(u).
\label{eom_fk}
\end{eqnarray}
\end{subequations}

\noindent
In the equation (\ref{eom_f0}),
avoiding the singularity at the horizon,
we fix the function $F_0(u)$ as a constant $C$ which is determined later,
\begin{equation}
F_0(u)=C.
\end{equation}
Using this solution, the equation (\ref{eom_fk}) leads
\begin{equation}
F'_{k^2}(u)=C\frac{z^2_H}{4uf(u)}
\!\int_1^u\!\!\dd u'\ \ee^{\frac{1}{2}\big(\frac{z_H}{z_Q}\big)^2\!u'}
=C\frac{z^2_H}{4uf(u)}2\Big(\frac{z_Q}{z_H}\Big)^2
\bigg(\ee^{\frac{1}{2}\big(\frac{z_H}{z_Q}\big)^2\!u}
-\ee^{\frac{1}{2}\big(\frac{z_H}{z_Q}\big)^2}\bigg),
\end{equation}
where we have imposed the regularity at the horizon.
We insert the solutions above into the equation (\ref{eom_at})
and obtain
\begin{equation}
\frac{z_H^2}{4uf(u)}
\Big(k^2 A_t(u)+\omega kA_z(u)\Big)
=X'(u)=k^2 F'_{k^2}(u)+{\cal O}(\omega).
\label{eom_at_exp}
\end{equation}
Taking the definition of the quark number susceptibility (\ref{suss})
into account, the equation (\ref{eom_at_exp}) is enough to
determine $A_t(u)$,
\begin{eqnarray}
A_t(u)
&=&
2C\Big(\frac{z_Q}{z_H}\Big)^2
\bigg(\ee^{\frac{1}{2}\big(\frac{z_H}{z_Q}\big)^2\!u}
-\ee^{\frac{1}{2}\big(\frac{z_H}{z_Q}\big)^2}\bigg)
\nonumber
\\
&=&
\frac{A_t^0}{1-\ee^{\frac{1}{2}\big(\frac{z_H}{z_Q}\big)^2}}
\bigg(
\ee^{\frac{1}{2}\big(\frac{z_H}{z_Q}\big)^2\!u}
-\ee^{\frac{1}{2}\big(\frac{z_H}{z_Q}\big)^2}
\bigg),
\end{eqnarray}
where we have defined the boundary variable as
$A_t^0\equiv A_t(u)|_{u\to 0}$ to fix the constant $C$.

Let us evaluate the quark number susceptibility.
The on-shell action to be needed is estimated as
\begin{equation}
S=\frac{1}{g_5^2 z_H^2}
\!\int\!\frac{\dd^4k}{(2\pi)^4}
\ \ee^{-\frac{1}{2}\big(\frac{z_H}{z_Q}\big)^2\!u}
A_t(-k)A'_t(k)\bigg|_{u=1}^{u=0}.
\end{equation}
Following the procedure given in Appendix A, we can read off
the two point function $G_{t \ t}(\omega, k)$.
We then conclude that the quark number susceptibility is
given as
\begin{equation}
\chi_q
=
\frac{1}{g_5^2z_Q^2
\bigg(\ee^{\frac{1}{2}\big(\frac{z_H}{z_Q}\big)^2}-1\bigg)}.
\label{qsuss_hydro}
\end{equation}

\begin{figure}[!ht]
\begin{center}
\includegraphics[angle=0, width=0.55\textwidth]{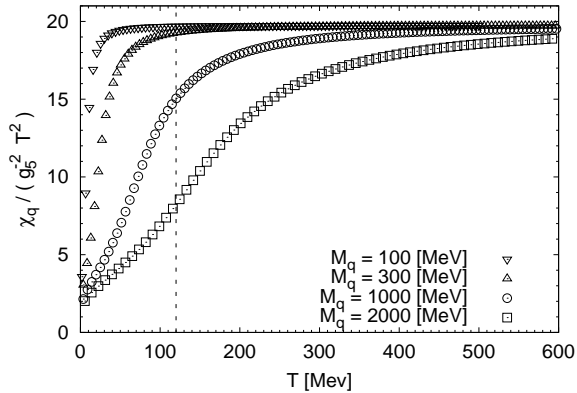}
\caption{The quark number susceptibility as a function of
the temperature for various quark masses.
The vertical dashed line denotes
the deconfinement temperature in the hard wall model.
}
\label{fig:fig1}
\end{center}
\end{figure}

In Fig.\ref{fig:fig1},
we plot $\chi_q/T^2$ as a function of $T$ with varying the
quark mass $M_q$\footnote{
We are interested in the qualitative tendency of the 
mass dependence to the quark number susceptibility. 
For the illustration purpose, 
we vary $M_q$ from 100 to 2000 MeV.  
}. 
Note here that the quark mass in our model
can be different from that in QCD (or in the chiral Lagrangian)
by a constant factor.
For instance, in the hard wall model the factor is $\sqrt{3}$,
which is obtained by matching the scalar two-point correlators
calculated from QCD and the hard wall
model~\cite{PR2}: $M_q^{\rm ChPT}=M_q/3$.
If we assume that this is valid in our case, we have to divide
the quark mass in our model by the factor $\sqrt{3}$ when we compare
our results with those from QCD or chiral perturbation theory (ChPT).
In Fig.\ref{fig:fig1}, the vertical dashed line is for a typical
deconfinement temperature with no quark mass effect
in the hard wall model~\cite{Herzog}.
If we assume that the deconfinement temperature in our approach
is roughly equal to that in the hard wall model\footnote{
We should consider the Hawking-Page type transition~\cite{hp, hpw} 
in the current setup.
Since the analysis for this confinement/deconfinement in \cite{KMS} needs
to be improved as we explained,
we here take the hard wall model as the closest existing one.
},
our results on the left of this line are just for illustration and
might be replaced by those with the thermal AdS background as
in \cite{jkls, KMSTT}.
As expected the finite quark mass suppresses the quark number
susceptibility.
We find that at high temperatures $T\ge 600$ MeV the quark number
susceptibility of light quarks and heavy quarks are almost equal.
This indicates that the heavy quark like charm contribution to
thermodynamics of a QCD-like system may start to become significant
at temperatures $T \sim 600$ MeV.
Though it might not mean much, we compare our number $T\sim 600$ with
that from lattice QCD.
In \cite{PHV}, it is shown that the fluctuations of the charm quark
number is comparable to those of the light quark sector at high
temperatures and claim that at temperatures $T\sim 800$ MeV
the contribution of charm quark to QCD thermodynamics is significant.

Before closing this section,
we briefly demonstrate the thermodynamic analysis on this
system.
In the background (\ref{metric_dads_02}),
we turn on the temporal component of gauge field which
could introduce the finite density in the boundary gauge theory.
We work in the grand canonical ensemble.
In the AdS/CFT correspondence,
we could identify the grand potential $\Omega(T, \mu)$
as
\begin{equation}
\Omega=T S_{\rm on\mbox{-}shell},
\label{grand_potential}
\end{equation}
where $S_{\rm on\mbox{-}shell}$ is the 5D bulk on-shell action
evaluated in Euclidean spacetime.
As in the usual imaginary time analysis,
we analytically continue to Euclidean spacetime and
compactify the time direction with the period $1/T$.
Assuming that the gauge field  depends only on the radial coordinate
$u$, the equation of motion leads
\begin{equation}
A'_t(u)=-d\ \ee^{\frac{1}{2}\big(\frac{z_H}{z_Q}\big)^2\!u},
\end{equation}
where the integration constant $d\ (>0)$ will be proportional
to the quark number density.
Then the grand potential (\ref{grand_potential}) can be
estimated as
\begin{equation}
\Omega
=
\frac{d^2 V_3}{g_5^2z_H^2}\!\int_1^0\!\dd u \
\ee^{\frac{1}{2}\big(\frac{z_H}{z_Q}\big)^2\!u}
=\frac{2d^2V_3z_Q^2}{g_5^2z_H^4}
\Big(1-\ee^{\frac{1}{2}\big(\frac{z_H}{z_Q}\big)^2}\Big),
\end{equation}
where $V_3$ is 3D volume.
The chemical potential is defined by
\begin{equation}
\mu=\!\int_1^0\!\dd u\ A'_t(u)
=-\frac{2dz_Q^2}{z_H^2}\Big(1-\ee^{\frac{1}{2}
\big(\frac{z_H}{z_Q}\big)^2}\Big).
\end{equation}
Using these relations, we obtain
\begin{equation}
\Omega=\frac{V_3\mu^2}{2g_5^2z_Q^2
\bigg(1-\ee^{\frac{1}{2}\big(\frac{z_H}{z_Q}\big)^2}\bigg)}.
\end{equation}
Since the quark number density is defined by
\begin{equation}
n_q=-\frac{1}{V_3}\frac{\del\Omega}{\del\mu}
\ \Big(=\frac{2}{g_5^2z_H^2}d\Big),
\end{equation}
one can easily confirm that the grand potential
exactly gives the quark number susceptibility
obtained in (\ref{qsuss_hydro}).
As it should be, the hydrodynamic and thermodynamic analysis
give the same result.

\section{Quark number susceptibility in D3/D7 system \label{SecD3D7}}

In this section,
we study effects of the chemical potential and the
quark mass for the quark number susceptibility at finite temperature
in a D3/D7 system.
Following \cite{D3D7density, kmmmt},
we consider the ``black hole embedding'' of the probe D7-branes
in the D3-brane background in which we could introduce
the finite quark mass together with the chemical potential.
In this section
we follow the procedure developed in \cite{kmmmt} and
then extract relevant parts to
discuss the quark number susceptibility.

We start by introducing $N_c$ D3-branes.
The AdS/CFT correspondence provides
the dual descriptions of ${\cal N}=4$ $SU(N_c)$ SYM as
the type IIB string theory on AdS$_5\times$S$^5$ with
the identifications $L^4/l_s^4=2g^2_{\rm YM} N_c=2\lambda$ and
$g_s=g^2_{\rm YM}/(2\pi)$.
Here $L$, $l_s=\sqrt{\alpha'}$ and $\lambda$ are the radius
in AdS$_5$ and S$^5$, the string length and 't Hooft coupling,
respectively.
Taking the large $N_c$ and the large 't Hooft coupling limit,
the string description reduces to that for the supergravity.
The finite temperature system in the deconfined phase
can be realized by using a corresponding black hole geometry in the
Euclidean signature:
\begin{equation}
(\dd s)^2=
\frac{u^2}{L^2}
\Big(f_0(\dd t)^2
+(\dd\vec{x})^2
\Big)
+\frac{L^2}{u^2}
\Big(
\frac{(\dd u)^2}{f_0}
+u^2\dd\Omega_5^2
\Big),
\label{D3geom_0}
\end{equation}
with
$$
f_0(u)=1-\frac{u_0^4}{u^4},
$$
where $u_0$ implies the location of horizon.
The periodic thermal identification
$t\simeq t+1/T$ leads
the Hawking temperature
\begin{equation}
T=\frac{u_0}{\pi L^2}.
\end{equation}
Following the paper \cite{mmt},
we introduce new coordinate $\varrho$ through
\begin{equation}
\varrho^2=u^2+\sqrt{u^4-u_0^4},
\end{equation}
and rewrite the metric (\ref{D3geom_0}) as
\begin{eqnarray}
(\dd s)^2
&=&
\frac{1}{2}\frac{\varrho^2}{L^2}
\left(
\frac{f^2}{\tilde f}(\dd t)^2
+\tilde{f}(\dd\vec{x})^2
\right)
+
\frac{L^2}{\varrho^2}
\Big((\dd\varrho)^2+\varrho^2 \dd\Omega_5^2\Big),
\label{D3geom}
\end{eqnarray}
with
$$
f(\varrho) = 1-\frac{u_0^4}{\varrho^4}, \qquad
\tilde{f}(\varrho)=1+\frac{u_0^4}{\varrho^4}.
$$
We work with the dimensionless coordinate
$\rho\equiv\varrho/u_0$ in the lest of the paper.
We refer to the horizon as $\rho=1$ and the AdS boundary
as $\rho\ (=\sqrt{2}u/u_0)\to\infty$.

In order to introduce the fundamental matters,
we may consider flavor branes~\cite{kk},
which provide flavor gauge fields in the world volume.
Here we put $N_f$ D7-branes with
the following intersection
\begin{equation}
\begin{array}{ccccccccccc}
& 0 & 1 & 2 & 3 & 4 & 5 & 6 & 7 & 8 & \ 9 \\
\mbox{D3:} & \times & \times & \times & \times & & & & & & \\
\mbox{D7:} & \times & \times & \times & \times & \times & \times
& \times & \times & & \\
\end{array}
\end{equation}
where D7-branes are wrapping on S$^3$ of S$^5$.
The additional degrees of freedom which we would like to discuss
are generated by open string
oscillations between D3 and D7-branes described by
${\cal N}=2$ hypermultiplet.
The fermions in this multiplet could be interpreted as the
analog of quarks in QCD.
Taking the probe approximation $N_f\ll N_c$,
we could consider the overall dynamics of $N_f$ D7-branes
which can be described by the DBI action in the
10D background (\ref{D3geom}),
\begin{equation}
S=N_fT_7\!\int\!\dd^8\sigma\ \ee^{-\phi}
\sqrt{\ \det(g_{mn}+2\pi\alpha'F_{mn})},
\label{dbi}
\end{equation}
where
$g_{mn}(\sigma)$ and $F_{mn}(\sigma)$ are the induced metric and
the field strength of the world volume gauge field,
respectively.
The tension of D7-branes and the dilaton field are given
by $T_7=1/((2\pi)^7l_s^8)$ and $\ee^\phi=g_s$, respectively.
It is convenient to divide the transverse 6D part to
D3-branes in (\ref{D3geom})
into two parts i.e.\ 4D and 2D whose coordinates are given by
spherical $(r, \Omega_3)$ and polar $(R, \varphi)$ coordinates,
respectively,
\begin{eqnarray}
(\dd\rho)^2 + \rho^2\dd\Omega_5^2
&=&
(\dd r)^2 + r^2\dd\Omega_3^2 + (\dd\!R)^2+R^2(\dd\varphi)^2
\nonumber
\\
&=&
(\dd\rho)^2 + \rho^2
\Big(
(\dd\theta)^2 + \sin^2\theta  \, \dd\Omega_3^2
+ \cos^2\theta \, (\dd\varphi)^2
\Big),
\label{met2}
\end{eqnarray}
where $r=\rho\sin\theta$,
$R=\rho\cos\theta$ with $0\le\theta\le\pi/2$
and $\rho^2=r^2+R^2$.
By construction,
we consider the case where the world volume coordinates of D7-branes
are given in the static
gauge i.e.\ $\sigma^m\equiv(t, \vec{x}, \rho, \Omega_3)$.
Due to the symmetries for the translation in $(t, \vec{x})$ and the
rotation in $(\rho, \Omega_3)$, the embedding of the D7-branes
could only depend on the radial coordinate $\rho$.
Since the rotational symmetry in $(R, \varphi)$ allows
to be $\varphi=0$, the embedding might be characterized by
$\chi(\rho)\equiv\cos\theta$ through $\theta(\rho)$ which is the angle
between two spaces $(r, \Omega_3)$ and $(R, \varphi)$.
The asymptotic value of the distance between D3 and D7-branes
which is given by $R(\rho)$ for large $\rho$ provides the quark mass
$M_q$.
The induced metric on the D7-branes is given by
\begin{eqnarray}
(\dd s)^2_{\rm D7}
&=&
L^2\Bigg\{
\frac{\pi^2T^2}{2}\rho^2
\left(
\frac{f^2}{\tilde f}(\dd t)^2
+ \tilde{f} (\dd\vec{x})^2
\right)
+
\frac{1}{\rho^2}
\left(
\frac{1-\chi^2 + \rho^2 \dot{\chi}^2 }{1-\chi^2}
\right)
(\dd\rho)^2
\nonumber
\\
&&
\hspace*{7mm}
+
\big(1-\chi^2\big)\ \dd\Omega_3^2
\Bigg\},
\quad
\label{D3geom-induced}
\end{eqnarray}
where the dot stands for the derivative with respective to $\rho$.
We also introduce the non-dynamical temporal component of the
gauge field $A_t(\rho)$ which leads the chemical potential and
the density at the AdS boundary.

By using the induced metric (\ref{D3geom-induced}) and the gauge
potential $A_t(\rho)$,
the DBI action (\ref{dbi}) now becomes
\begin{equation}
S_0=\frac{\lambda N_cN_fT^3}{32}V_3
\!\int\!\dd\rho\
\rho^3\tilde{f}(1-\chi^2)
\sqrt{f^2(1-\chi^2+\rho^2\dot\chi^2)
-2\tilde{f}(1-\chi^2)\dot{\widetilde{A}_t^2}},
\label{dimensionless}
\end{equation}
where we have defined
$\widetilde{A}_t(\rho)\equiv 2\pi\alpha'A_t(\rho)/u_0$.
Since there exist no $\widetilde{A}_t(\rho)$ terms in the action,
the equation of motion for $\widetilde{A}_t(\rho)$ can be
reduced to the following form with an integration
constant $\widetilde{d}$,
\begin{equation}
\widetilde{d}\equiv
\frac{\rho^3\tilde{f}^2(1-\chi^2)^2\dot{\widetilde{A}_t}}
{2\sqrt{f^2(1-\chi^2+\rho^2\dot{\chi}^2)-2\tilde{f}(1-\chi^2)
\dot{\widetilde{A}^2_t}}}.
\label{eom_ta}
\end{equation}
The equation of motion for $\chi(\rho)$ is given as
\begin{eqnarray}
0
&=&
\del_\rho
\Bigg\{
\frac{\rho^5f\tilde{f}(1-\chi^2)\dot{\chi}}
{\sqrt{1-\chi^2+\rho^2\dot{\chi}^2}}
\Bigg(1+\frac{8\widetilde{d}^2}{\rho^6\tilde{f}^3(1-\chi^2)^3}\Bigg)^{1/2}
\Bigg\}
\nonumber
\\
&&+\frac{\rho^3f\tilde{f}\chi}{\sqrt{1-\chi^2+\rho^2\dot{\chi}^2}}
\Bigg\{
\Big(3(1-\chi^2)+2\rho^2\dot{\chi}^2\Big)
\Bigg(1+\frac{8\widetilde{d}^2}{\rho^6\tilde{f}^3(1-\chi^2)^3}\Bigg)^{1/2}
\nonumber
\\
&&
\hspace*{34mm}
-\frac{24\widetilde{d}^2(1-\chi^2+\rho^2\dot{\chi}^2)}
{\rho^6\tilde{f}^3(1-\chi^2)^3}
\Bigg(1+\frac{8\widetilde{d}^2}{\rho^6\tilde{f}^3(1-\chi^2)^3}\Bigg)^{-1/2}
\Bigg\},
\quad\quad
\label{eom_chi}
\end{eqnarray}
where we have eliminated the gauge field $\widetilde{A}_t(\rho)$ by
using the relation (\ref{eom_ta}).
Near the boundary, asymptotic solutions of the equations of motion
(\ref{eom_ta}) and (\ref{eom_chi}) behave as
\begin{subequations}
\begin{eqnarray}
\widetilde{A}_t(\rho)
&=&
\widetilde{\mu}-\frac{\widetilde{d}}{\rho^2}+\cdots,
\label{aspt_a}
\\
\chi(\rho)
&=&
\frac{m}{\rho}+\frac{c}{\rho^3}+\cdots.
\label{aspt_chi}
\end{eqnarray}
\end{subequations}

\noindent
The chemical potential $\mu$ can be defined as the boundary
value of $A_t(\rho)$,
while the quark mass $M_q$ can be estimated through the asymptotic
value of the separation of D3 and
D7-branes i.e.\ $M_q=\rho\chi(\rho)/(2\pi\alpha')$ at the AdS boundary.
Taking the rescaling of the gauge field $\widetilde{A}_t$ and
the coordinate $\rho$ into account, the integration constants
$\widetilde{\mu}$ and $m$ can be related with these values,
\begin{subequations}
\begin{eqnarray}
\widetilde{\mu}
&=&
\frac{2\pi\alpha'}{u_0}\mu
=\sqrt{\frac{2}{\lambda}}\frac{\mu}{T},
\\
m
&=&
2\pi\alpha'\frac{\sqrt{2}}{u_0}M_q
=\frac{2}{\sqrt{\lambda}}\frac{M_q}{T}.
\label{mMq}
\end{eqnarray}
\end{subequations}

\noindent
As we will estimate below\footnote{
In the dimension three operator,
we neglect contributions from squarks in the hypermultiplet.},
the remaining constants $\widetilde{d}$ and $c$ would be proportional
to the the quark number density $n_q$ and the quark
condensate $\langle\bar{\psi}\psi\rangle$, respectively.

We have to solve the equations of motion (\ref{eom_ta})
and (\ref{eom_chi}) numerically to determine the embedding
and the property of the gauge field.
Here we restrict to the black hole embedding in which
the D7-branes touch the horizon,
since this might be thermodynamically favored configuration
in the system with finite density.
We impose boundary conditions at the horizon
as $\dot{\chi}(1)=0$ and $\widetilde{A}_t(1)=0$ to remove singularities
and $\chi(1)=\chi_0$.
We fix $m$ and $\widetilde{\mu}$ which depend on $\chi_0$ and
$\widetilde{d}$ by matching the numerical solutions
with the asymptotic forms at the boundary.

We now consider the on-shell action which is related with
the partition function $Z$ of the field theory in the context of
AdS/CFT correspondence:
\begin{equation}
Z=\ee^{-S_{\rm on\mbox{-}shell}}.
\end{equation}
However the on-shell action in this system contains UV divergences.
It is well-known that one can prepare local boundary counter terms
for probe D-branes in AdS spacetime by applying the holographic
renormalization~\cite{kos}.
In the D3/D7 system, taking the asymptotic solution (\ref{aspt_chi})
into account, the relevant boundary counter terms take the
form~\cite{mmt},
\begin{equation}
S_{\rm ct}
=\frac{\lambda N_cN_fT^3}{32}V_3
\Bigg\{
-\frac{1}{4}
\Big((\rho_{\rm max}^2-m^2)^2-4mc\Big)
\Bigg\},
\end{equation}
where $\rho_{\rm max}$ is the cut-off for UV divergences
which may go to infinity after precise calculations.
It should be noticed that there exist finite contributions
in the counter terms.
Together with these counter terms,
we could obtain the regularized action
\begin{eqnarray}
S
&=&
S_0 + S_{\rm ct}
\nonumber
\\
&=&
\frac{\lambda N_cN_fT^3}{32}V_3
\Bigg\{
\int_1^{\infty}\!\!
\dd\rho
\bigg(\rho^3\tilde{f}(1-\chi^2)
\sqrt{f^2(1-\chi^2+\rho^2\dot{\chi}^2)
-2\tilde{f}(1-\chi^2)\dot{\widetilde{A}_t^2}}
\nonumber
\\
&&
\hspace*{39mm}
-\rho^3+m^2\rho\bigg)
-\frac{1}{4}\Big((m^2-1)^2-4mc\Big)
\Bigg\}.
\label{reg_action}
\end{eqnarray}
Since we are interested in the black hole embedding,
the integration supports from the horizon
to the AdS boundary.
Evaluating the on-shell action for (\ref{reg_action})
which is reduced to boundary values through the equation of motion,
we could observe that the quark condensate is proportional to
the integration constant $c$,
\begin{equation}
\langle\bar{\psi}\psi\rangle
=-\frac{T}{V_3}\frac{\del}{\del M_q}\log Z
=-\frac{1}{8}\sqrt{\lambda}N_cN_fT^3c,
\label{qc}
\end{equation}
where we have used the asymptotic solutions (\ref{aspt_a}) and
(\ref{aspt_chi}) and the relation (\ref{mMq}).
As we did in the previous section,
we could identify the grand potential $\Omega(\mu, T)$ as
\begin{equation}
\Omega=-T\log Z=TS_{\rm on{\mbox{-}}shell}.
\end{equation}
The quark number density can be also calculated through
the on-shell evaluation,
\begin{equation}
n_q=-\frac{1}{V_3}\frac{\del\Omega}{\del\mu}
=\frac{1}{4}\sqrt{\frac{\lambda}{2}}N_cN_fT^3\widetilde{d}.\label{D3D7nq}
\end{equation}

\begin{figure}[!ht]
\begin{center}
\includegraphics[angle=0, width=0.5\textwidth]{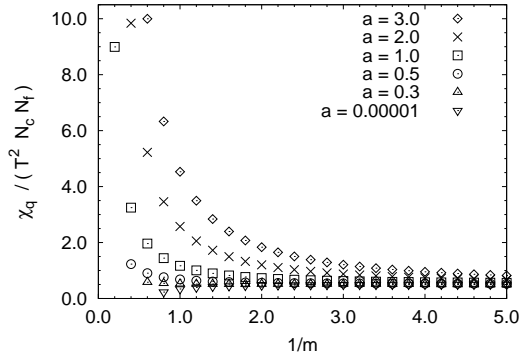}
\caption{The quark number susceptibility in the D3/D7 model
as a function of $1/m$.
Here $a= \widetilde\mu/m=\mu/(\sqrt{2}M_q)$.} \label{fig:fig2}
\end{center}
\end{figure}

Let us now calculate the quark number susceptibility
numerically using (\ref{def_qns}) and (\ref{D3D7nq}).
In Fig.\ref{fig:fig2}, we plot our results
with $a= \widetilde\mu/m\sim \mu/M_q$.
Since $1/m\sim T/M_q$, we can consider the horizontal axis
of the figure as the temperature with the choice of
$M_q$ fixed.
For very small $a$ ($\ll 1$) or $M_q\gg \mu/\sqrt{2}$,
the quark mass is dominating the quark number susceptibility
and suppresses it at low temperature,
while in the opposite case the chemical potential enhances it.
Note that the confinement/deconfinement transition temperature 
in the D3/D7 model is zero, since there is no dimensionful parameter 
which could characterize the critical temperature as in the Hawking-Page
transition~\cite{hp, hpw}. 
Indeed, within the large $N_c$ approximation, 
one can observe the entropy is zero for $T=0$, while for given finite 
$T$ that is proportional to $N_c^2T^3$ whose $N_c^2$ dependence 
indicates the deconfinement of the color degrees of freedom~\cite{hpw}.  
In addition, if one would like to consider mesons which correspond to 
the fluctuations of D7 flavor branes, one could observe their discrete 
mass spectra at $T=0$~\cite{kmmw}. 
Another cautionary remark is that in the QCD phase diagram,  
there exists a CEP  at a single value of the chemical potential,
while Fig.2 may imply multiple points in the sense that the quark number 
susceptibility is diverging
for many different values of the chemical potential with fixed quark mass. 
At this moment, we have no clear understanding how to remove 
multiple CEPs. We may hope that some large $N_c$ corrections 
could resolve this problem. 
Without resolving this issue, we may not be able to address
single CEP in the QCD phase diagram within D3/D7 model.

Now we discuss implications of our findings in QCD at high temperature.
We first spell out some limitations of our approach.
As well known, a generic problem with this kind of holographic
QCD is that the results from holographic QCD are mostly
large $N_c$ leading ones.
Though there are some studies to include the corrections,
most of them include some subset of the corrections,
not all of them in a consistent way.
Apart from this generic plague, the D3/D7 model we used here shows
that deconfinement temperature $T_c$ is zero.
In addition, we assumed an exact isospin symmetry to use
the abelian DBI action for the $N_f$ probe branes,
and so our approach could not study $N_f$-dependent nature of
QCD phase transition.
Unlike QCD, the temperature $T$, the chemical potential $\mu$,
and the quark mass $M_q$ are not all independent,
and only two of them or ratios of them are independent,
which is the relic of conformal nature of our background metric.
With the above-mentioned in mind, we comment on QCD,
more precisely a QCD-like system, at finite temperature
based on Fig.\ref{fig:fig2}.

For small  $a$ ($\ll 1$) or $M_q\gg \mu/\sqrt{2}$,
the quark number susceptibility does not show any diverging behavior
as the temperature approaches $T_c$,
indicating that if the quark mass is big enough compared
to the value of the quark chemical potential,
the transition to deconfined phase is a smooth crossover
rather than first or second order phase transition.
In the opposite case $a>1$, it shows a rapid jump-up,
coming close to $T_c$ from high temperature.
This implies that as long as the quark chemical potential is
a few times larger than the quark mass, the transition would be
second order.
In Appendix \ref{AppB}, we calculate the chiral susceptibility
in the D3/D7 model, see Fig.\ref{fig:fig3}.
We observe that it shows a similar behavior with the quark number
susceptibility in Fig.\ref{fig:fig2}.

\section{Summary}

We studied the effect of a finite quark mass on the quark number
susceptibility in the framework of holographic QCD.
We first considered the bottom-up model with the deformed AdS black
hole that includes the back-reaction of the quark mass.
As expected the finite quark mass suppressed the quark number
susceptibility.
We observed that at high temperatures $T\ge 600$ MeV the quark
number susceptibility of light quarks and heavy quarks
are almost equal, indicating  that the heavy quark like charm contribution
to thermodynamics of a QCD-like system may start to become
significant at such temperatures.

We then moved to the D3/D7 model to calculate the quark number
susceptibility at finite temperature with a finite quark chemical
potential.
We studied the competition between the quark chemical potential,
which enhances the quark number susceptibility, and the quark mass
that suppresses the susceptibility.
We observed that depending on the relative values of the quark mass
and the quark chemical potential, the quark number susceptibility
shows diverging or converging behavior.
With a caution that our approach is still not that close to
a realistic system like quark-gluon plasma, we discussed physical
implication of our results in a QCD-like system at the end of
section \ref{SecD3D7}.

We also calculated the chiral susceptibility in D3/D7 model
in Appendix \ref{AppB} to support the observation made with
the quark number susceptibility.

\section*{Acknowledgements}

YK thanks Y. Seo for helpful discussions. KK thanks S.H. Lee for 
letting him know the importance of the quark mass effect on the
quark number susceptibility. ST would like to thank T. Misumi and
I.J. Shin for useful discussions and K. Anagnostopoulos for
technical supports. ST used the cluster system in National Technical
University of Athens and B-Factory Computer System of KEK. YK and TT
acknowledge the Max Planck Society~(MPG), the Korea Ministry of
Education, Science and Technology~(MEST), Gyeongsangbuk-Do and
Pohang City for the support of the Independent Junior Research Group
at the Asia Pacific Center for Theoretical Physics~(APCTP). The work
of KK was supported by the Korean BK21 Program and Korea Research
Foundation(KRF-2006-C00011).

\appendix

\section{Minkowskian correlators in AdS/CFT correspondence}

In this appendix,
we briefly summarize the prescription for the Minkowskian correlator
in AdS/CFT correspondence.
We here follow the prescription proposed in~\cite{Son:2002sd}.
We work on the following 5D background,
\begin{equation}
 (\dd s)^2 = g_{\mu\nu}\dd x^\mu\dd x^\nu + g_{uu}(\dd u)^2,
\end{equation}
where $x^\mu$ and $u$ are the 4D and radial coordinates,
respectively.
We refer the boundary at $u=0$ and the horizon at $u=1$.
Let us consider a solution of an equation of motion in
this 5D background.
Suppose a solution of an equation of motion is given by
\begin{equation}
\phi(u,x) =
\!\int\!\frac{\dd^4 k}{(2\pi)^4}\ \mbox{e}^{ikx}f_k(u)\phi^0(k),
\end{equation}
where $f_k(u)$ is normalized such that $f_k(0)=1$ at the boundary.
After putting the equation of
motion back into the action,
the on shell action might be reduced to surface terms
\begin{equation}
S[\phi^0]
=\!\int\!\frac{{\rm d}^4 k}{(2\pi)^4}
\phi^0(-k){\cal G}(k, u)\phi^0(k)
\bigg|_{u=1}^{u=0}.
\label{on_shell_action}
\end{equation}
Here, the function $\mathcal G(k,u)$ can be written
in terms of $f_{\pm k}(u)$ and $\partial_u f_{\pm k}(u)$.
Accommodating Gubser-Klebanov-Polyakov/Witten
relation~\cite{gkp,w} to Minkowski spacetime,
Son and Starinets proposed the formula to get the retarded Green functions,
\begin{equation}
G(k)
=
2{\cal G}(k, u)
\bigg|_{u=0},
\label{green_function}
\end{equation}
where the incoming boundary condition at the horizon is imposed.
In this paper, we consider correlators of $U(1)$
currents $J_\mu(x)$, where $J_\mu (x)$ is the vector current of
quark field or quark number current.
Now we define the precise form of the retarded Green
functions which we discuss in this paper:
\begin{equation}
G_{\mu\ \nu}(k)
=-i\!\int\!\dd^4x \ {\rm e}^{-ikx}\theta(t)
\big\langle
[J_\mu (x), \ J_\nu (0)]
\big\rangle.
\label{cc}
\end{equation}
%

\section{Chiral susceptibility in D3/D7 system\label{AppB}}

In this appendix, we consider the chiral susceptibility.
Note that in D3/D7 system, we have only $U(1)$ axial symmetry.
Chiral susceptibility $\chi_c$ is one of the important
observables in terms of chiral symmetry restoration:
\begin{eqnarray}
\chi_c
\equiv
\frac{\del \langle \bar{\psi} \psi \rangle }{\del M_q}.
\end{eqnarray}
In the D3/D7 system,
the quark condensate $\langle\bar{\psi}\psi\rangle$
is given by (\ref{qc}).
Note that our chiral condensate $\langle \bar{\psi} \psi \rangle$
is defined to be positive since $c$ is negative,
while it is negative in ordinary QCD.
Reading out the numerical value of $c$, we calculate the chiral
susceptibility.
\begin{figure}[!ht]
\begin{center}
\includegraphics[angle=0, width=0.5\textwidth]{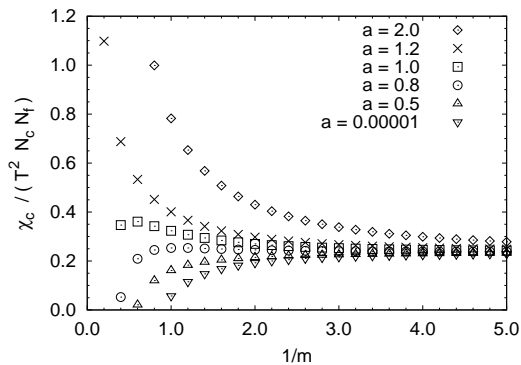}
\caption{ The chiral susceptibility as function of $1/m$.} \label{fig:fig3}
\end{center}
\end{figure}



\begin{thebibliography}{99}

\bibitem{Alltonetal}
C.R. Allton, M. Doring, S. Ejiri, S.J. Hands, O. Kaczmarek, F. Karsch,
E. Laermann and K. Redlich,
Phys. Rev. {\bf D71} (2005) 054508,
{\tt[arXiv:hep-lat/0501030]}.

\bibitem{Ejiri}
S. Ejiri, C.R. Allton, M. Doring, S.J. Hands,  O. Kaczmarek, F. Karsch,
E. Laermann and K. Redlich,
Nucl. Phys. {\bf A774} (2006) 837,
{\tt[arXiv:hep-ph/0509361]};
\\
R.V. Gavai and S. Gupta,
Phys. Rev. {\bf D71} (2005) 114014,
{\tt[arXiv:hep-lat/0412035]}.

\bibitem{CEP}
S. Gupta,
{\tt[arXiv:0909.4630[nucl-ex]]};
\\
R. Gavai and S. Gupta,
PoS LAT2005 (2006) 160,
{\tt[arXiv:hep-lat/0509151]};
\\
M.A. Stephanov,
Prog. Theor. Phys. Suppl. {\bf 153} (2004) 139,
Int. J. Mod. Phys. {\bf A20} (2005) 4387,
{\tt[arXiv:hep-ph/0402115]}.

\bibitem{Mc87}
L. McLerran, Phys. Rev. {\bf D36} (1987) 3291.

\bibitem{Gottlieb:1987ac}
S.A. Gottlieb, W. Liu, D. Toussaint, R.L. Renken and R.L. Sugar,
Phys.\ Rev.\ Lett.\ {\bf 59} (1987) 2247.

\bibitem{Kunihiro:91}
T. Kunihiro,
Phys.\ Lett.\ {\bf B271} (1991) 395.

\bibitem{CMT}
P. Chakraborty, M.G. Mustafa and M.H. Thoma,
Eur.\ Phys.\ J.\ {\bf C23} (2002) 591, \\
{\tt[arXiv:hep-ph/0111022]};
\\
J.P. Blaizot, E. Iancu and A. Rebhan,
Phys.\ Lett.\  {\bf B523} (2001) 143, \\
{\tt [arXiv:hep-ph/0110369]};
Eur.\ Phys.\ J.\  {\bf C27} (2003) 433,
{\tt [arXiv:hep-ph/0206280]}.

\bibitem{HI}
Y. Hatta and T. Ikeda,
Phys. Rev. {\bf D67} (2003) 014028,
{\tt[arXiv:hep-ph/0210284]}.

\bibitem{HKRS}
M. Harada, Y. Kim, M. Rho and C. Sasaki,
Nucl. Phys. {\bf A727} (2003) 437, \\
{\tt[arXiv:hep-ph/0207012]}.

\bibitem{Sasaki}
C. Sasaki, B. Friman, and K. Redlich,
Phys. Rev. {\bf D75} (2007) 074013, \\
{\tt[arXiv:hep-ph/0611147]}.

\bibitem{HM2010}
N. Haque and M.G. Mustafa,
{\tt [arXiv:1007.2076[hep-ph]]}.

\bibitem{GLTRS}
S. Gottlieb, W. Liu, D. Toussaint, R.L. Renken and R.L. Sugar,
Phys.\ Rev.\ Lett.\ {\bf 59} (1987) 2247.

\bibitem{GLTRS88}
S. Gottlieb, W. Liu, D. Toussaint, R.L. Renken and R.L. Sugar,
Phys. Rev. {\bf D38} (1988) 2888.

\bibitem{GPS}
R.V. Gavai, J. Potvin and  S. Sanielevici,
Phys. Rev. {\bf D40} (1989) 2743.

\bibitem{GKetal}
S.A. Gottlieb {\it et al.},
Phys. Rev. {\bf D55} (1997) 6852,
{\tt[arXiv:hep-lat/9612020]}.

\bibitem{EKR}
S. Ejiri, F. Karsch and K. Redlich,
Phys. Lett. {\bf B633} (2006) 275,
{\tt[arXiv:hep-ph/0509051]}.

\bibitem{jkls}
K. Jo, Y. Kim, H.K. Lee and S.-J. Sin,
JHEP {\bf 0811} (2008) 040, \\
{\tt [arXiv:0810.0063[hep-ph]]}.

\bibitem{KMSTT}
Y. Kim, Y. Matsuo, W. Sim, S. Takeuchi and T. Tsukioka,
JHEP {\bf 1005} (2010) 038, \\
{\tt[arXiv:1001.5343[hep-th]]}.

\bibitem{SZ}
A. Stoffers and I. Zahed, Phys.Rev. {\bf D83} (2011) 055016,
{\tt[arXiv:1009.4428[hep-th]]}.

\bibitem{Maldacena}
J.M. Maldacena,
Adv.\ Theor.\ Math.\ Phys.\  {\bf 2} (1998) 231,
[Int.\ J.\ Theor.\ Phys.\  {\bf 38} (1999) 1113],
{\tt [arXiv:hep-th/9711200]}.

\bibitem{gkp}
S.S. Gubser, I.R. Klebanov and A.M. Polyakov,
Phys.\ Lett.\ {\bf B428} (1998) 105, \\
{\tt [arXiv:hep-th/9802109]}.

\bibitem{w}
E. Witten,
Adv.\ Theor.\ Math.\ Phys.\ {\bf 2} (1998) 253,
{\tt [arXiv:hep-th/9802150]}.

\bibitem{Son:2002sd}
D.T. Son and A.O. Starinets,
JHEP {\bf 0209} (2002) 042,
{\tt[arXiv:hep-th/0205051]}.

\bibitem{pss}
G. Policastro, D.T. Son and A.O. Starinets,
JHEP {\bf 0209} (2002) 043, \\
{\tt[arXiv:hep-th/0205052]}.

\bibitem{kss}
P. Kovtun, D.T. Son and A.O. Starinets,
JHEP {\bf 0310} (2003) 064, \\
{\tt[arXiv:hep-th/0309213]};
Phys. Rev. Lett. {\bf 94} (2005) 111601,
{\tt[arXiv:hep-th/0405231]}.

\bibitem{LS}
M. Laine and Y. Schroder,
Phys. Rev. {\bf D73} (2006) 085009,
{\tt[arXiv:hep-ph/0603048]}.

\bibitem{KMS}
Y. Kim,  T. Misumi and I.J. Shin,
{\tt[arXiv:0911.3205[hep-ph]]}.

\bibitem{D3D7density}
S. Nakamura, Y. Seo, S.-J. Sin and K.P. Yogendran,
J. Korean Phys. Soc. {\bf 52} (2008) 1734, \\
{\tt[arXiv:hep-th/0611021]};
Prog. Theor. Phys. {\bf 120} (2008) 51, \\
{\tt [arXiv:0708.2818[hep-th]]}.

\bibitem{kmmmt}
S. Kobayashi, D. Mateos, S. Matsuura, R.C. Myers and R.M. Thomson,
JHEP {\bf 0702} (2007) 016,
{\tt[arXiv:hep-th/0611099]}.

\bibitem{hp}
S.W. Hawking and D.N. Page, Commun. Math. Phys. {\bf 87} (1983) 577. 

\bibitem{hpw}
E. Witten, Adv. Theor. Math. Phys. {\bf 2} (1998) 505, 
{\tt[arXiv:hep-th/9803131]}. 

\bibitem{PR2}
L. Da Rold and A. Pomarol,
JHEP {\bf 0601} (2006) 157,
{\tt [arXiv:hep-ph/0510268]}.

\bibitem{Herzog}
C.P. Herzog,
Phys. Rev. Lett. {\bf 98} (2007) 091601,
{\tt [arXiv:hep-th/0608151]}.

\bibitem{PHV}
P. Petreczky, P. Hegde and A. Velytsky,
PoS LAT2009 (2009) 159, \\
{\tt[arXiv:0911.0196[hep-lat]]}.

\bibitem{kk}
A. Karch and E. Katz,
JHEP {\bf 0206} (2002) 043,
{\tt[arXiv:hep-th/0205236]}; \\
J. Babington, J. Erdmenger, N.J. Evans, Z. Guralnik and I. Kirsch,
Phys. Rev. {\bf D69} (2004) 066007,
{\tt [arXiv:hep-th/0306018]}.

\bibitem{kos}
A. Karch, A. O'Bannon and K. Skenderis,
JHEP {\bf 0604} (2006) 015, \\
{\tt[arXiv:hep-th/0512125]}.

\bibitem{mmt}
D. Mateos, R.C. Myers and R.M. Thomson,
Phys. Rev. Lett. {\bf 97} (2006) 091601, 
\\
{\tt[arXiv:hep-th/0605046]};
JHEP {\bf 0705} (2007) 067,
{\tt[arXiv:hep-th/0701132]}.

\bibitem{kmmw}
M. Kruczenski, D. Mateos, R.C. Myers and D.J. Winters, 
JHEP {\bf 0307} (2003) 049, \\
{\tt[arXiv:hep-th/0304032]}. 

\end{thebibliography}
\end{document}